\documentclass[runningheads]{llncs}
\usepackage[cmex10]{amsmath}
\usepackage{amssymb}
\usepackage{./mathSymbols}
\usepackage{graphicx}
\usepackage[pdftex,dvipsnames]{xcolor}  % Colored text etc.
\usepackage{xargs}   % Use more than one optional parameter in a new
\usepackage{algorithmicx}
\usepackage[ruled,noend,linesnumbered]{algorithm2e}
\SetKwRepeat{Do}{do}{while}
\usepackage{multirow}
\usepackage[]{todonotes}

\usepackage{makecell}

\usepackage{caption}
\usepackage{subcaption}
\usepackage{scalefnt}
\usetikzlibrary{positioning,shapes,arrows,fit}
\usepackage{bm,times}
\usepackage[firstpage]{draftwatermark}\SetWatermarkText{\hspace*{5in}\raisebox{9.2in}{\includegraphics[scale=0.1]{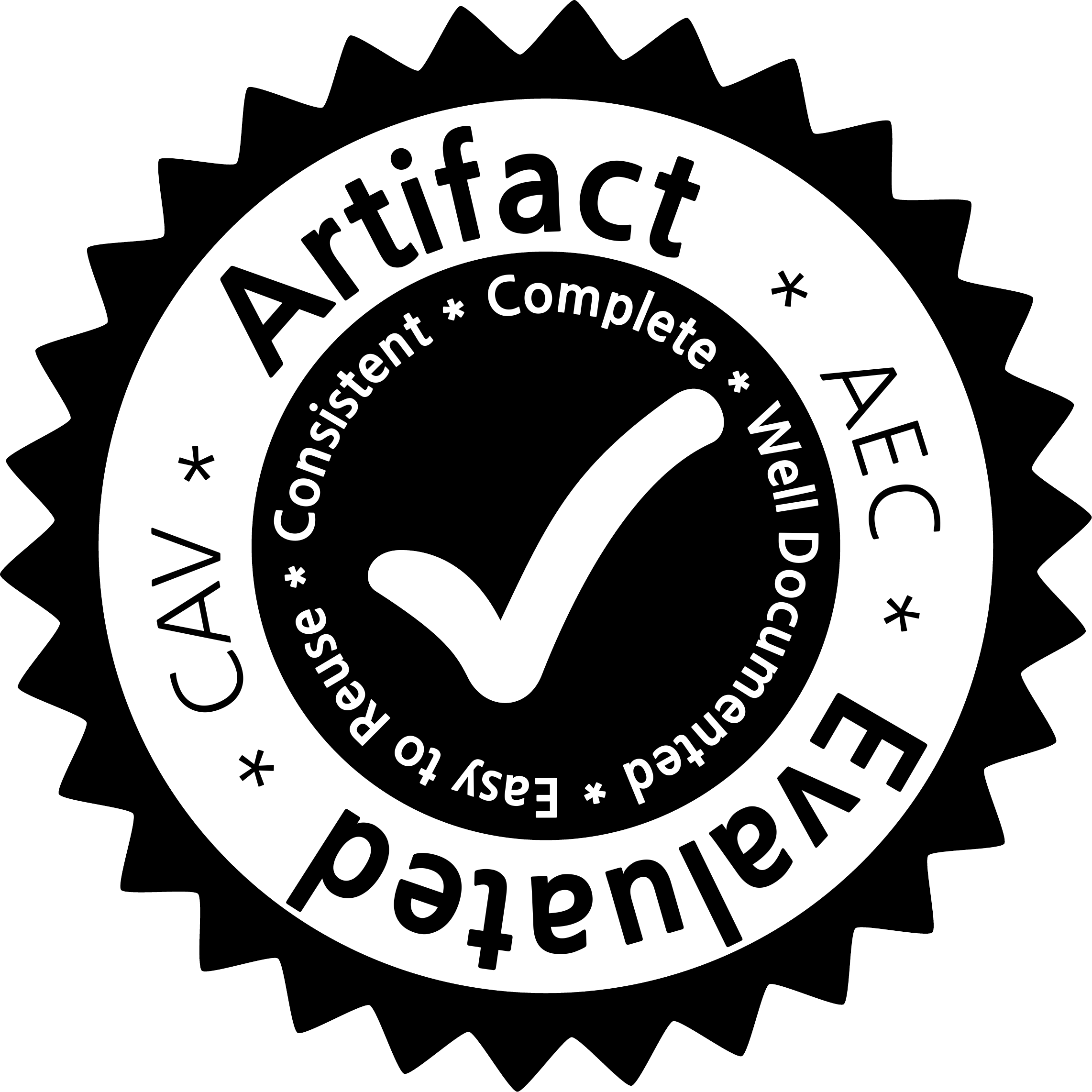}}}\SetWatermarkAngle{0}

\newcommandx{\tn}[2][1=]{\todo[inline,
  linecolor=LimeGreen,backgroundcolor=LimeGreen!40,bordercolor=LimeGreen,#1]{(Thakur)
    #2}}
\newcommandx{\cjm}[2][1=]{\todo[inline,
  linecolor=YellowOrangeGreen,backgroundcolor=YellowOrange!40,bordercolor=YellowOrange,#1]{(Chris)
    #2}}
\newcommandx{\ckm}[2][1=]{\todo[inline,
  linecolor=MidnightBlue,backgroundcolor=MidnightBlue!40,bordercolor=MidnightBlue,#1]{(Curtis)
    #2}}
\newcommandx{\hz}[2][1=]{\todo[inline,
  linecolor=Orchid,backgroundcolor=Orchid!40,bordercolor=Orchid,#1]{(Hao)
    #2}}
\newcommandx{\zz}[2][1=]{\todo[inline,
  linecolor=RawSienna,backgroundcolor=RawSienna!40,bordercolor=RawSienna,#1]{(Zhen)
    #2}}
\newcommandx{\all}[2][1=]{\todo[inline,
  linecolor=red,backgroundcolor=red!80,bordercolor=red,#1]{(All) #2}}

\begin{document}
%
% \title{Contribution Title\thanks{Supported by organization x.}}
\title{STAMINA: STochastic Approximate Model-checker for INfinite-state Analysis 
%\\ {\bf OR}
%\textcolor{red}{STAMINA: A New Infinite-State CTMC Model Checker}
%\\
}
%Improved Approximation Techniques for Probabilistic Analysis of
%  Continuous-Time Markov Chains}
%
\titlerunning{STAMINA: A New Infinite-state CTMC Model-checker}


\author{Thakur Neupane\inst{1}\orcidID{0000-0002-1870-4079} \and
  Chris J. Myers\inst{2}\orcidID{0000-0002-8762-8444} \and \\
  Curtis Madsen\inst{3}\orcidID{0000-0002-0254-0364} \and
  Hao Zheng\inst{4}\orcidID{0000-0002-8627-0591} \and \\
  Zhen Zhang\inst{1}\orcidID{0000-0002-8269-9489}
  }
\authorrunning{T. Neupane et al.}
\institute{Utah State University, Logan, UT USA, 
  \email{thakur.neupane@aggiemail.usu.edu, zhen.zhang@usu.edu}\\
\and University of Utah, Salt Lake City, UT USA, \email{myers@ece.utah.edu}\\  
\and Boston University, Boston, MA USA, \email{ckmadsen@bu.edu}\\
\and University of South Florida, Tampa, FL USA, \email{haozheng@usf.edu}}
\maketitle              % typeset the header of the contribution
\begin{abstract}
Stochastic model checking is a technique for analyzing systems that
possess probabilistic characteristics.  However, its scalability is
limited as probabilistic models of real-world applications typically
have very large or infinite state space. This paper presents a new
infinite state CTMC model checker, STAMINA, with improved scalability.
It uses a novel state space approximation method to reduce large and
possibly infinite state CTMC models to finite state representations
that are amenable to existing stochastic model checkers.  It is
integrated with a new property-guided state expansion approach that
improves the analysis accuracy.  Demonstration of the tool on several
benchmark examples shows promising results in terms of analysis
efficiency and accuracy compared with a state-of-the-art CTMC model
checker that deploys a similar approximation method. 

\end{abstract}
\section{Introduction}

Stochastic model checking is a formal method that designers and engineers can use to determine the likelihood of \emph{safety} and \emph{liveness} properties.
%if their system exhibits \emph{safety} (i.e., the system never enters an error state), \emph{liveness} (i.e., the system eventually enters an operating state), \emph{sequentiality} (i.e., the system repeatedly enters a sequence of states), or arbitrarily combinations of these properties.
Checking properties using numerical model checking techniques requires enumerating the state space of the system to determine the probability that the system is in any given state at a desired time~\cite{Kwiatkowska2007}.
Real-world applications often have very large or even infinite state spaces.  

Numerous state representation, reduction, and approximation methods have been proposed. Symbolic model checking based on \emph{multi-terminal binary decision diagrams} (MTBDDs)~\cite{Parker2002} has achieved success in representing large \emph{Markov Decision Process} (MDP) models with a few distinct probabilistic choices at each state, e.g., the shared coin protocol~\cite{Aspnes1990}. MTBDDs, however, are often inefficient for models with many different and distinct probability/rate values due to the inefficient representation of solution vectors. \emph{Continuous-time Markov chain} (CTMC) models, whose state transition rate is a function of state variables, generally contain many distinct rate values. As a result, symbolic model checkers can run out of memory while verifying a typical CTMC model with as few as 73,000 states~\cite{Parker2002}. State reduction techniques, such as bisimulation minimization~\cite{Katoen2007,Fisler1999,Fisler2002}, abstraction~\cite{Madsen2014,Katoen2007,Fecher2006,Hermanns2008}, symmetry reduction~\cite{Kwiatkowska2006,Donaldson2006}, and partial order reduction~\cite{Groesser2005} have been mainly extended to discrete-time, finite-state probabilistic systems. The three-valued abstraction~\cite{Katoen2007} can reduce large, finite-state CTMCs. It may, however, provide inconclusive verification results due to abstraction.

To the best of our knowledge, only a few tools can analyze infinite-state probabilistic models, namely, STAR~\cite{Lapin2011} and INFAMY~\cite{Hahn2009}. The STAR tool primarily analyzes biochemical reaction networks. It approximates solutions to the \emph{chemical master equation} (CME) using the \emph{method of conditional moments} (MCM)~\cite{Hasenauer2014} that combines moment-based and state-based representations of probability distributions. This hybrid approach represents species with low concentrations using a discrete stochastic description and numerically integrates a small master equation using the fourth order Runge-Kutta method over a small time interval~\cite{Andreychenko2011}; and solves a system of conditional moment equations for higher concentration species, conditioned on the low concentration species. This method has been optimized to drop unlikely states and add likely states on-the-fly. STAR relies on a well-structured underlying Markov process with small sensitivity on the transient distribution. Also, it mainly reports state reachability probabilities, instead of checking a given probabilistic property. INFAMY is a truncation-based approach that explores the model's state space up to a certain finite depth $k$. The truncated state space still grows exponentially with respect to exploration depth. Starting from the initial state, breadth-first state search is performed 
%layer by layer 
up to a certain finite depth. The error probability computed during the model checking depends on the depth of state exploration. Therefore, higher exploration depth generally incurs lower error probability. 
%Error estimation methods implemented include \emph{finite state projection (FSP)}, \emph{uniform}, and \emph{layered} to maintain a small error probability. 
% Compared to INFAMY, our method allows different depth on each search apth, and only terminates a path if the corresponding estimated path probability falls below a user defined threshold.

This paper presents a new infinite-state stochastic model checker,  \emph{STochastic Approximate Model-checker for INfinite-state Analysis} (STAMINA). Our tool also takes a truncation-based approach.  In particular, it maintains a probability estimate of each path being explored in the state space, and when the currently explored path probability drops below a specified threshold, it halts exploration of this path. All transitions exiting this state are redirected to an absorbing state.  After all paths have been explored or truncated, transient Markov chain analysis is applied to determine the probability of a transient property of interest specified using \emph{Continuous Stochastic Logic} (CSL)~\cite{Aziz2000}.  The calculated probability forms a lower bound on the probability, while the upper bound also includes the probability of the absorbing state.  The actual probability of the CSL property is guaranteed to be within this range.  An initial version of our tool and preliminary results are reported in~\cite{Neupane2019}.  Since that paper, our tool has been tightly integrated within the PRISM model checker~\cite{Kwiatkowska2011} to improve performance, and we have also developed a new property-guided state expansion technique to expand the state space to tighten the reported probability range incrementally.  This paper reports our results, which 
%on this improved version of STAMINA. In particular, results 
show significant improvement on both efficiency and verification accuracy over several non-trivial case studies from various application domains. 

%%% Local Variables:
%%% mode: latex
%%% TeX-master: "main"
%%% End:

%\input{relatedWork}
\section{STAMINA}
\label{sec-methods}

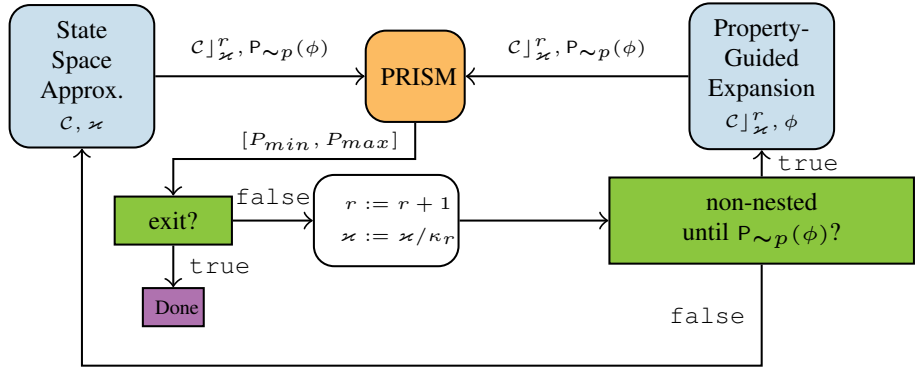
\begin{figure}[tbhp]
\tikzstyle{every node}=[font=\tiny]
\tikzstyle{stExp}=[draw, fill=MidnightBlue!15, text width=2.5em, 
text centered, minimum height=2.5em, rounded corners]
% \tikzstyle{framework}=[draw, fill=MidnightBlue!15, text width=7em, 
%     text centered, minimum height=9em, rounded corners]
\tikzstyle{prism} = [stExp, text width=1.5em, fill=YellowOrange!70,
minimum height=2em, rounded corners]
\tikzstyle{dec} = [draw, fill=LimeGreen, text width=1.9em, minimum height=0.18em, text centered]
\def\blockdist{2}
\def\edgedist{2}
\resizebox{\textwidth}{!}{%
\begin{tikzpicture}
  \node (stApprox) [stExp] {State Space Approx.\\\scalebox{0.7} {$\ctmcShort, \stTermToler$}};%, \epsilon, N$}};
  \path (stApprox.east) +(0.9*\blockdist, 0) node (prism) [prism] {PRISM};

  \path (stApprox.east) +(2.1*\blockdist,0) node (propGuided) [stExp] {Property-Guided\\Expansion\\\scalebox{0.7} {$\ctmcTruncShort{\stTermToler}^{r}, \cslPath$}};

  \node (termCond) [dec, below of=stApprox, xshift=2em] {exit?};
  \path [draw, ->] (stApprox.east) -- node [above] {\scalebox{0.7} {$\ctmcTruncShort{\stTermToler}^{r}, \csl$}} (prism.west);
  \path [draw, ->] (prism.south) -- node [left] {\scalebox{0.7} {$[P_{min}, P_{max}]$}} ++(0, -0.25) -| (termCond);
  \path (termCond) +(0, -0.3*\blockdist) node (done) [draw, rectangle, scale=0.8, fill=Orchid, text width=1em] {Done};
  \path [draw, ->] (termCond.south) -- node [right] {\texttt{true}} (done);
  \node (update) [draw, right of=termCond, xshift=1.5em, text width=2.5em, text centered, minimum height=2em, rounded corners] {\scalebox{0.7} {
      $\begin{aligned}
        r &:= r+1\\
        \stTermToler &:= \stTermToler/\kappa_r
      \end{aligned}$}};
  \path [draw, ->] (termCond.east) -- node [above] {\texttt{false}} (update);
%   \path (update) +(0.65*\blockdist, 0) node (nestedCSL) [dec]{nested \scalebox{0.7}{$\csl$}?};
%   \path [draw, ->] (update) -- (nestedCSL);
%   \path [draw, ->] (nestedCSL.south) -- node [left] {\texttt{true}} ++(0, -0.5) -| (stApprox.south);
  % \path [draw, ->, to path={-| (\tikztotarget)}] (nestedCSL.south) edge (a); 
%   \path (nestedCSL) +(0.67*\blockdist, 0) node (until) [dec]{\scalebox{0.8}{$\cslPath = \text{until}$?}};
  \path (update) +(1.3*\blockdist, 0) node (until) [dec,text width=6em]{non-nested\\until \scalebox{0.8}{$\csl$}?};
  \path [draw, ->] (update) -- (until);
%   \path [draw, ->] (nestedCSL.east) -- node [above] {\texttt{false}} (until.west);
  \path [draw, ->] (until) -- node [right] {\texttt{true}} (propGuided);
  \path [draw, ->] (until.south) -- node [left] {\texttt{false}} ++(0, -0.7) -| (stApprox.south);
  \path [draw, ->] (propGuided.west) -- node [above] {\scalebox{0.7} {$\ctmcTruncShort{\stTermToler}^{r}, \csl$}} (prism.east);
\end{tikzpicture}
}
\caption{Architecture of STAMINA.} %The exiting condition is $p \not\in [P_{min}, P_{max}] \lor (P_{max} - P_{min}) < \epsilon \lor r > N$.}
\label{fig:stamina-architecture}
\end{figure}

Figure~\ref{fig:stamina-architecture} presents the architecture of STAMINA. Based on a user-specified probability threshold $\stTermToler$ (kappa), it first constructs a finite-state CTMC model $\ctmcTruncShort{\stTermToler}$ from the original infinite-state CTMC model $\ctmcShort$ using the state space approximation method presented in Section~\ref{sec:state-space-approx}. $\ctmcTruncShort{\stTermToler}$ is then checked using the PRISM explicit-state model checker against a given CSL property $\csl$, where $\thicksim \in \{<, \; >, \; \leqslant, \; \geqslant\}$ and $p \in [0, 1]$ (for cases where it is desired that a predicate be true within a certain probability bound) or
% $\thicksim \!\! p \equiv = ?$ 
$\cslExact$ (for cases where it is desired that the exact probability of the predicate being true be calculated). Lower- and upper-bound probabilities that $\cslPath$ holds, namely, $P_{min}$ and $P_{max}$, are then obtained, and their difference, i.e., $(P_{max} - P_{min})$, is the probability accumulated in the absorbing state $\stAbsorb$ which abstracts all the states not included in the current state space. If $p \in [P_{min}, P_{max}]$, it is not known whether $\csl$ holds. If exact probability is of interest and the probability range is larger than the user-defined precision $\epsilon$, i.e., $(P_{max} - P_{min}) > \epsilon$, then the method does not give a meaningful result.

For an inconclusive verification result from the previous step, 
STAMINA applies a property-guided approach, described in Section~\ref{sec:property-based-exploration},  to further expand $\ctmcTruncShort{\stTermToler}$, provided $\csl$ is a non-nested ``until'' formula; otherwise, it uses the previous method to expand the state space. % Section~\ref{sec:property-based-exploration} describes this method. 
Note that $\stTermToler$ also drops by the reduction factor $\kappa_r$ to enable states that were previously ignored due to a low probability estimate to be included in the current state expansion. The expanded CTMC model $\ctmcTruncShort{\stTermToler}$ is then checked to obtain a new probability bound $[P_{min}, P_{max}]$. This iterative process repeats until one of the following conditions holds: (1) the target probability $p$ falls outside the probability bound $[P_{min},P_{max}]$, (2) the probability bound is sufficiently small, i.e,  $(P_{max} - P_{min}) < \epsilon$, or (3) a maximal number of iterations $N$ has been reached ($r \geqslant N$). 

\subsection{State Space Approximation}
\label{sec:state-space-approx}
The state space approximation method~\cite{Neupane2019} truncates the state space based on a user-specified reachability threshold $\stTermToler$. During state exploration, the reachability-value function, $\stTermCurShort : \stSet{} \rightarrow \posReal$, estimates the probability of reaching a state on-the-fly, and is compared against $\stTermToler$ to determine whether the state search should terminate. Only states with a higher reachability-value than the reachability threshold are explored further. 

\begin{figure*}[htbp]
\centering
  \begin{subfigure}[t]{0.2\textwidth}
  \centering
    \begin{tikzpicture}
    \pgfmathsetmacro{\dist}{1.5}

      \begin{scope}[every node/.style={circle,thick,draw,align=center,minimum size=1cm}]
            \node (S0) at (0,       0) {$\st_0$\\$1.0$};
      \end{scope}

        \draw [->,>=stealth,black,thick] (-1,0) -- node [text width=1cm,midway,above]{Start} (-0.5,0);

    \end{tikzpicture}
    \caption{}
    \label{fig-approx-init}
  \end{subfigure}%
  \begin{subfigure}[t]{0.2\textwidth}
  \centering
        \begin{tikzpicture}
    \pgfmathsetmacro{\dist}{1.5}
      \begin{scope}[every node/.style={circle,thick,draw,align=center,minimum size=1cm}]
            \node (S0) at (0,       0) {$\st_0$\\$0.0$};
            \node (S1) at (1*\dist, 0) {$\st_1$\\$0.8$};
            \node (S4) at (0,       1*\dist) {$\st_4$\\$0.2$};
      \end{scope}
      \begin{scope}[>=stealth,every edge/.style={draw=black,very thick}]
    
        \path [->] (S0) edge[bend left=20] node[above]{$0.8$} (S1);
        \path [->] (S0) edge[bend left=20] node[left]{$0.2$} (S4);
        
    \end{scope}

        \draw [->,>=stealth,black,thick] (-1,0) -- node [text width=1cm,midway,above]{Start} (-0.5,0);

    \end{tikzpicture}
    \caption{}
    \label{fig-approx-step1}
  \end{subfigure}%
  \begin{subfigure}[t]{0.6\textwidth}
  \centering
    \begin{tikzpicture}
    \pgfmathsetmacro{\dist}{1.5}
      \begin{scope}[every node/.style={circle,thick,draw,align=center,minimum size=1cm}]
            \node (S0) at (0,       0) {$\st_0$\\$0.2$};
            \node (S1) at (1*\dist, 0) {$\st_1$\\$0.0$};
            \node (S2) at (2*\dist, 0) {$\st_2$\\$0.15$};
            \node (S4) at (0,       1*\dist) {$\st_4$\\$0.0$};
            \node (S5) at (1*\dist, 1*\dist) {$\st_5$\\$0.65$};
      \end{scope}
      \begin{scope}[>=stealth,every edge/.style={draw=black,very thick}]
        \path [->] (S0) edge[bend left=20] (S1);
        \path [->] (S0) edge[bend left=20] (S4);
        \path [->] (S1) edge[bend left=20] (S2);
        \path [->] (S1) edge[bend left=20] (S0);
        \path [->] (S1) edge[bend left=20] (S5);
        % \path [->] (S4) edge[bend left=20] (S5);
        \path [->] (S4) edge[bend left=20] (S0);
    \end{scope}

        \draw [->,>=stealth,black,thick] (-1,0) -- node [text width=1cm,midway,above]{Start} (-0.5,0);

    \end{tikzpicture}
    \caption{}
    \label{fig-approx-step2}
  \end{subfigure}%
  \\
  \begin{subfigure}[t]{0.4\textwidth}
  \centering
    \begin{tikzpicture}
    \pgfmathsetmacro{\dist}{1.5}
      \begin{scope}[every node/.style={circle,thick,draw,align=center,minimum size=1cm}]
            \node (S0) at (0,       0) {$\st_0$\\$0.2$};
            \node (S1) at (1*\dist, 0) {$\st_1$\\$0.23$};
            \node (S2) at (2*\dist, 0) {$\st_2$\\$0.12$};
            \node (S4) at (0,       1*\dist) {$\st_4$\\$0.1$};
            \node (S5) at (1*\dist, 1*\dist) {$\st_5$\\$0.0$};
            \node (S6) at (2*\dist, 1*\dist) {$\st_6$\\$0.18$};
            \node (S9) at (1*\dist, 2*\dist) {$\st_9$\\$0.17$};
      \end{scope}
      \begin{scope}[>=stealth,every edge/.style={draw=black,very thick}]
        \path [->] (S0) edge[bend left=20] (S1);
        \path [->] (S0) edge[bend left=20] (S4);
        \path [->] (S1) edge[bend left=20] (S2);
        \path [->] (S1) edge[bend left=20] (S0);
        \path [->] (S1) edge[bend left=20] (S5);
        \path [->] (S2) edge[bend left=20] (S1);
        % \path [->] (S4) edge[bend left=20] (S5);
        \path [->] (S4) edge[bend left=20] (S0);
        \path [->] (S5) edge[bend left=20] (S6);
        \path [->] (S5) edge[bend left=20] (S9);
        \path [->] (S5) edge[bend left=20] (S4);
        \path [->] (S5) edge[bend left=20] (S1);
    \end{scope}

        \draw [->,>=stealth,black,thick] (-1,0) -- node [text width=1cm,midway,above]{Start} (-0.5,0);

    \end{tikzpicture}
    \caption{}
    \label{fig-approx-step3}
  \end{subfigure}%
  \begin{subfigure}[t]{0.6\textwidth}
  \centering
    \begin{tikzpicture}
    \pgfmathsetmacro{\dist}{1.5}
      \begin{scope}[every node/.style={circle,thick,draw,align=center,minimum size=1cm}]
            \node (S0) at (0,       0) {$\st_0$\\$0.2$};
            \node (S1) at (1*\dist, 0) {$\st_1$\\$0.23$};
            \node (S2) at (2*\dist, 0) {$\st_2$\\$0.12$};
            \node (S4) at (0,       1*\dist) {$\st_4$\\$0.1$};
            \node (S5) at (1*\dist, 1*\dist) {$\st_5$\\$0.13$};
            \node (S6) at (2*\dist, 1*\dist) {$\st_6$\\$0.12$};
            \node (S9) at (1*\dist, 2*\dist) {$\st_9$\\$0.1$};
            \node[fill=gray] (Sabs) at (2*\dist,2.1*\dist) {$\stAbsorb$};
      \end{scope}
      \begin{scope}[>=stealth,every edge/.style={draw=black,very thick}]
        \path [->] (S0) edge[bend left=20] (S1);
        \path [->] (S0) edge[bend left=20] (S4);
        \path [->] (S1) edge[bend left=20] (S2);
        \path [->] (S1) edge[bend left=20] (S0);
        \path [->] (S1) edge[bend left=20] (S5);
        \path [->] (S2) edge[bend left=20] (S1);
        \path [->] (S2) edge[bend left=20] (S6);
        \path [->] (S4) edge[bend left=20] (S5);
        \path [->] (S4) edge[bend left=20] (S0);
        \path [->] (S5) edge[bend left=20] (S9);
        \path [->] (S5) edge[bend left=20] (S6);
        \path [->] (S5) edge[bend left=20] (S4);
        \path [->] (S5) edge[bend left=20] (S1);
        \path [->] (S6) edge[bend left=20] (S5);
        \path [->] (S6) edge[bend left=20] (S2);
        \path [->] (S9) edge[bend left=20] (S5);
        \path [->] (S2) edge[bend right=60] (Sabs);
        \path [->] (S6) edge[bend right=00] (Sabs);
        \path [->] (S9) edge[bend right=00] (Sabs);
        \path [->] (S4) edge[bend left=60] (Sabs);
    \end{scope}

        \draw [->,>=stealth,black,thick] (-1,0) -- node [text width=1cm,midway,above]{Start} (-0.5,0);

    \end{tikzpicture}
    \caption{}
    \label{fig-approx-step4}
  \end{subfigure}
  \caption{State space approximation.}\label{fig-state-space-approx}
\end{figure*}
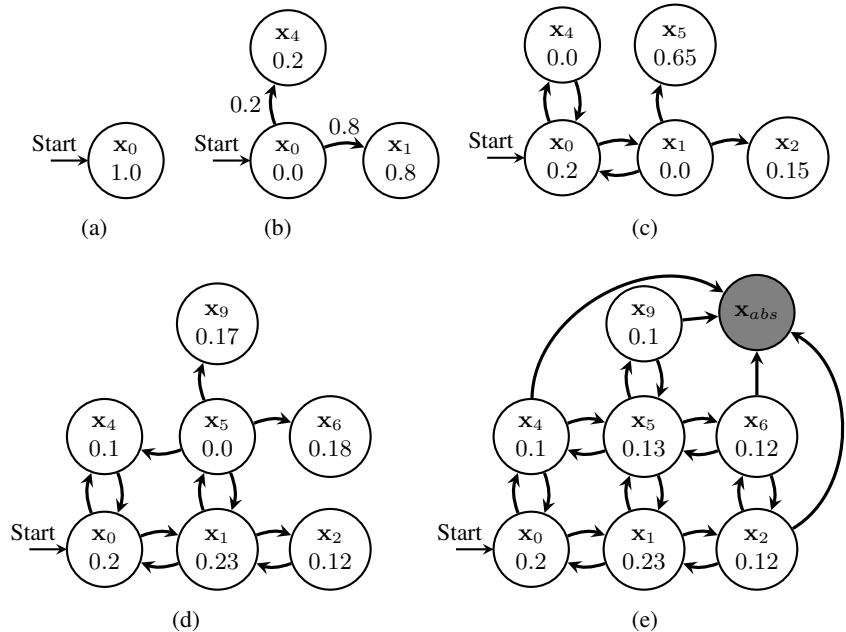

Figure~\ref{fig-state-space-approx} illustrates the standard \emph{breadth first search} (BFS) state exploration for reachability threshold $\stTermToler=0.25$. It starts from the initial state whose reachability-value i.e., $\stTermCurShort(\initSt)$, is initialized to $1.0$ as shown in Figure~\ref{fig-approx-init}. In the first step, two new states $\st_1$ and $\st_4$ are generated and their reachability-values are $0.8$ and $0.2$, respectively, as shown in Figure~\ref{fig-approx-step1}. The reachability-value in $\initSt$ is distributed to its successor states, based on the probability of outgoing transitions from $\initSt$ to its successor state. For the next step, only state $\st_1$ is scheduled for exploration because $\stTermCurShort(\st_1) \geq \stTermToler$. Note that the transition from $\st_4$ to $\st_0$ is executed  because $\st_0$ is already in the explored set. Expanding $\st_1$ leads to two new states, namely $\st_2$ and $\st_5$ as shown in Figure~\ref{fig-approx-step2}, from which only $\st_5$ is scheduled for further exploration. This leads to the generation of $\st_6$ and $\st_9$ shown in Figure~\ref{fig-approx-step3}. State exploration terminates after Figure~\ref{fig-approx-step4} since both newly generated states have reachability-values less than $0.25$. States $\st_2$, $\st_4$, $\st_6$ and $\st_9$ are marked as terminal states. During state exploration, the reachability-value update is performed every time a new incoming path is added to a state because a new incoming path can add its contribution to the state, potentially bringing the reachability-value above $\stTermToler$, which in turn changes a terminal state to be non-terminal. When the truncated CTMC model $\ctmcTruncShort{\stTermToler}$ is analyzed, it introduces some error in the probability value of the property under verification, because of leakage the probability (i.e., cumulative path probabilities of reaching states not included in the explored state space) during the CTMC analysis. To account for probability loss, an abstract absorbing state $\stAbsorb$ is created as the sole successor state for all terminal states on each truncated path. Figure~\ref{fig-approx-step4} shows the addition of the absorbing state. 

\subsection{Property Based State Space Exploration}
\label{sec:property-based-exploration}

This paper introduces a property-guided state expansion method, in order to efficiently obtain a tightened probability bound. Since all non-nested CSL path formulas $\cslPath$ (except those containing the ``next'' operator) derive from the ``until'' formula, $\untilFull$, construction of the set of terminal states for further expansion boils down to eliminating states that are known to satisfy or dissatisfy $\untilUnbounded$. Given a state graph, a path starting from the initial state can never satisfy $\untilUnbounded$, if it includes a state satisfying $\lnot \preUntil \land \lnot \postUntil$.
Also, if a path includes a state satisfying $\postUntil$, satisfiability of $\untilUnbounded$ can be determined without further expanding this path beyond the first $\postUntil$-state.
% and a path can potentially satisfy $\untilUnbounded$ if it includes a state that satisfies $\postUntil$~\cite{Baier2000,Kwiatkowska2007}. 
% It should be noted that state search on a path terminates immediately if either $\preUntil$ or $\postUntil$ is encountered first.
Our property-guided state space expansion method identifies the path prefixes, from which satisfiability of $\untilUnbounded$ can be determined, and shortens them by making the last state of each prefix absorbing based on the satisfiability of $(\lnot \preUntil \vee \postUntil)$. Only the non-absorbing states whose path probability is greater than the state probability estimate threshold $\stTermToler$ are expanded further. For detailed algorithms of STAMINA, readers are encouraged to read~\cite{Neupane2019Thesis}.

\section{Results}

This section presents results on the following case studies to illustrate the accuracy and efficiency of STAMINA: a genetic toggle switch~\cite{Madsen2014,Neupane2019}; the following examples from the PRISM benchmark suite~\cite{Kwiatkowsa2012}: grid world robot, cyclic server polling system, and tandem queuing network; and the Jackson queuing network from INFAMY case studies~\cite{INFAMY_case_studies}. All case studies are evaluated on STAMINA and INFAMY, except the genetic toggle switch~\footnote{INFAMY generates arithmetic errors on the genetic toggle switch model.}. Experiments are performed on a 3.2 GHz AMD Debian Linux PC with six cores and 64 GB of RAM. For all experiments, the maximal number of iterations $N$ is set to $10$, and the reduction factor $\kappa_r$ is set to $1000$.  All experiments terminate due to $(P_{max} - P_{min}) < \epsilon$, where $\epsilon=10^{-3}$, before they reach $N$. STAMINA is freely available at: \url{https://github.com/formal-verification-research/stamina}.

We compare the runtime, state size, and verification results between STAMINA and INFAMY using the same precision $\epsilon=10^{-3}$. For all tables in this section, column $\stTermToler$ reports the probability estimate threshold used to terminate state generation in STAMINA. 
% Each model configuration occupies several rows in each table. The first row lists probabilities obtained from stochastic model checking on the state space constructed by property-agnostic state exploration~\cite{Neupane2019}, and successive rows list probabilities for the expanded state space obtained from the property-guided state expansion state exploration detailed in Section~\ref{sec-methods}. Currently, the property-guided state expansion state exploration only supports non-nested bounded transient properties. For other types of properties, it reverts back to the property-agnostic state expansion with reduced $\stTermToler$. 
The state space size is listed in column $|\sg{}|(K)$, where $K$ indicates one thousand states. Column $T(C/A)$ reports the state space construction (C) and analysis (A) time in seconds. For STAMINA, the total construction and analysis time is the cumulation of runtime for all $\stTermToler$ values for a model configuration. Columns $P_{min}$ and $P_{max}$ list the lower and upper probability bounds for the property under verification, and column $P$ lists the single probability value (within the precision $\epsilon$) reported by INFAMY. We select the best runtime reported by three configurations of INFAMY. The improvement in state size (column $|\sg{}|(X)$) and runtime (column $T(\%)$) are represented by the ratio of state count generated by INFAMY to that of STAMINA (higher is better) and percentage improvement in runtime (higher is better), respectively.

\textbf{Genetic toggle switch}. The genetic toggle switch circuit model has two inputs,  aTc and IPTG. It can be set to the OFF state by supplying it with aTc and can be set to the ON state by supplying it with IPTG \cite{Madsen2014}. Two important properties for a toggle switch circuit are the response time and the failure rate. The first experiments set IPTG to $100$ to measure the toggle switch's response time. It should be noted that the input value of 100 molecules of IPTG is chosen to ensure that the circuit switches to the ON state. The later experiments initialize IPTG to $0$ to compute the failure rate, i.e., the probability that the circuit changes state erroneously within a cell cycle of $2,100$ seconds (an approximation of the cell cycle in \emph{E. coli}~\cite{Zheng2016}). Initially, LacI is set to $60$ and TetR is set to $0$ for both experiments. The CSL property used for both experiments, $\prob_{=?} \; [true \; \until^{\leqslant 2100} \; (TetR>40 \; \land \; LacI<20)]$,  describes the probability of the circuit switching to the ON state within a cell cycle of $2,100$ seconds. The ON state is defined as LacI below 20 and TetR above 40 molecules.

\begin{table}[tbhp]
\begin{center}
% \vspace*{-6pt}
\caption{Verification results for genetic toggle switch.}
\label{tbl-resultsToggle}
% \vspace{2mm}
\begin{tabular}{|c|c|c|c|c|c|c|}
\hline
\multirow{2}{*}{$IPTG$}      & \multicolumn{6}{c|}{STAMINA} \\ \cline{2-7} 
                             & $\stTermToler$   & $|\sg{}|$   & $T(C/A)$  & $P_{min}$  & $P_{max}$  & Remark \\ \hline
\multirow{3}{*}{$100$}       & $10^{-3}$  & $ 1,127 $   & $0.15/0.67$  & $0.000000$ & $0.999671$ & \multirow{2}{*}{Property} \\ \cline{2-6} 
                             & $10^{-6}$  & $ 4,461 $   & $0.43/2.84$  & $0.966947$ & $0.992908$ & \multirow{2}{*}{Guided} \\ \cline{2-6} 
                             & $10^{-9}$  & $ 7,163 $   & $0.43/5.25$  & $0.991738$ & $0.991797$ & \\ \hline

\multirow{2}{*}{$100$}       & $10^{-6}$  & $ 5,171 $   & $0.17/1.90$   & $0.977942$ & $0.992850$ & \multirow{1}{*}{Property}  \\ \cline{2-6} 
                             & $10^{-9}$  & $ 8,908 $   & $0.18/3.74$  & $0.991739$ & $0.991797$ &  \multirow{1}{*}{Agnostic}\\ \hline

\multirow{3}{*}{$0$}         & $10^{-3}$  & $  182  $   & $0.05/0.07$  & $0.000000$ & $0.697500$ & \multirow{2}{*}{Property} \\ \cline{2-6} 
                             & $10^{-6}$  & $ 2,438 $   & $0.16/1.08$ & $0.008814$ & $0.060424$ & \multirow{2}{*}{Guided}\\ \cline{2-6} 
                             & $10^{-9}$  & $ 4,284 $   & $0.09/2.12$  & $0.013097$ & $0.013609$ & \\ \hline

\multirow{2}{*}{$0$}         & $10^{-6}$  & $ 2,446 $   & $0.16/1.05$    & $0.009169$ & $0.060420$ & \multirow{1}{*}{Property} \\ \cline{2-6} 
                             & $10^{-9}$  & $ 4,820 $   & $0.13/2.13$  & $0.013097$ & $0.013609$ & \multirow{1}{*}{Agnostic} \\ \hline

\end{tabular}
\end{center}
\vspace*{-15pt}
\end{table}

The property-agnostic state space is generated with the probability estimate threshold $\stTermToler=10^{-3}$. Table~\ref{tbl-resultsToggle} shows large probability bounds: $[0, 0.999671]$ for $\text{IPTG}=100$ and $[0, 0.6975]$ for $\text{IPTG} = 0$. It is obvious that they are significantly inaccurate w.r.t. the precision $\epsilon$ of $10^{-3}$. The $\stTermToler$ is then reduced to $10^{-6}$ and state generation switches to the property-guided state expansion mode, where the CSL property is used to guide state exploration, based on the previous state graph. Each state expansion step reduces the $\stTermToler$ value by a factor of $\kappa_r = 1000$. To measure the effectiveness of the property-guided state expansion approach, we compare state graphs generated with and without the property-guided state expansion, as indicated by the ``property agnostic'' and ``property guided'' rows in the table.
% we generated state graphs for corresponding values of $\stTermToler$ with property-guided state expansion turned off (indicated by ``property agnostic'' rows).
Property-guided state expansion reduces the size of the state space without losing the analysis precision for the same value of $\stTermToler$. Specifically, the state expansion approach reduces the state space by almost $20\%$ for the response rate experiment.

\textbf{Robot World}. This case study considers a robot moving in an $n$-by-$n$ grid and a janitor moving in a larger grid $Kn$-by-$Kn$, where the constant $K$ is used to significantly scale up the state space. The robot starts from the bottom left corner to reach the top right corner. The janitor moves around randomly. Either the robot or janitor can occupy one grid location at any given time. The robot also randomly communicates with the base station. The property of interest is the probability that the robot reaches the top right corner within $100$ time units while periodically communicating with the base station, encoded as $\prob_{=?} \; [ \; (\prob_{\geqslant 0.5} \; [ \; true \; \until^{\leqslant 7} \; communicate \;]) \; \until^{\leqslant 100} \; goal \;]$. 

Table~\ref{tbl-compare-stamina-infamy} provides a comparison of results for $K=1024,64$ and $n=64,32$. For smaller grid size i.e, 32-by-32, the robot can reach the goal with a high probability of $97.56\%$. Where as for a larger value of $n=64$ and $K=64$, the robot is not able to reach the goal with considerable probability. STAMINA generates precise results that are similar to INFAMY, while exploring less than half of states with shorter runtime.
% As we reduce the value of $\stTermToler$, the probability bound becomes tighter. 
% STAMINA, by selectively exploring the paths whose path probabilities are higher than the given threshold, can generate very precise results that are similar to INFAMY, while exploring nearly $3$ times fewer states. This also explains the significant shorter runtime used by STAMINA.

\begin{table}[tbhp]
\begin{center}
\caption{Comparison between STAMINA and INFAMY.}
\label{tbl-compare-stamina-infamy}
\begin{tabular}{|c|c||c|c|c|c||c|c|c||c|c|}
\hline
\multirow{2}{*}{Model}   & \multirow{2}{*}{Params} & \multicolumn{4}{c||}{STAMINA} & \multicolumn{3}{c||}{INFAMY}      &   \multicolumn{2}{c|}{Improvement} \\ \cline{3-11} 
                         &                             & \makecell{$|\sg{}|$\\$(K)$}  & \makecell{$T$\\$(C/A)$} & $P_{min}$   & $P_{max}$  & \makecell{$|\sg{}|$\\$(K)$} & \makecell{$T$\\$(C/A)$} & $P$               &   \makecell{$|\sg{}|$\\$(X)$}   &    \makecell{$T$\\$(\%)$}  \\ \hline \hline

\multirow{1}{*}{}        & \makecell{$32/$\\$64$}                     & $696$     & \makecell{$41/$\\$279$}  & $0.975$ & $0.975$ & $1,591$    & \makecell{$492/$\\$18$}    & $0.975$ & $2.3$  & $37.3$   \\ \cline{2-11} 
\multirow{1}{*}{Robot}   & \makecell{$32/$\\$1024$}                   & $696$     & \makecell{$41/$\\$258$}  & $0.975$ & $0.975$ & $1,591$    & \makecell{$501/$\\$18$}    & $0.975$ & $2.3$  & $42.4$   \\ \cline{2-11} 
\multirow{1}{*}{$(n/K)$} & \makecell{$64/$\\$64$}                     & $2,273$   & \makecell{$135/$\\$669$} & $1.46e$-$4$ & $1.68e$-$4$ & $5,088$    & \makecell{$1,625/$\\$53$}  & $1.5e$-$4$ & $2.2$ & $52.1$   \\ \cline{2-11} 
\multirow{1}{*}{}        & \makecell{$64/$\\$1024$}                   & $2,273$   & \makecell{$132/$\\$621$} & $1.46e$-$4$ & $1.68e$-$4$ & $5,088$    & \makecell{$1,625/$\\$53$}  & $1.5e$-$4$ & $2.2$ & $55.2$      \\ \hline \hline

\multirow{1}{*}{Jackson}        & \makecell{$4/$\\$5$}                & $201$     & \makecell{$22/$\\$51$}   & $0.865$ & $0.865$ & $635$       & \makecell{$109/$\\$5$}    & $0.865$ & $3.2$   & $36.1$   \\ \cline{2-11} 
\multirow{1}{*}{$(N/\lambda)$}  & \makecell{$5/$\\$5$}                & $2,539$   & \makecell{$990/$\\$996$} & $0.819$ & $0.819$ & $7,029$     & \makecell{$1668/$\\$108$} & $0.819$ & $2.8$   & $-11.8$  \\ \hline \hline

\multirow{1}{*}{Polling} & $12$                        & $19$      & \makecell{$3/$\\$21$}    & $1.0$  & $1.0$ & $74$        & \makecell{$1/$\\$2$}      & $1.0$ & $3.9$   & $-732.2$ \\ \cline{2-11} 
\multirow{1}{*}{$(N)$}   & $16$                        & $57$      & \makecell{$18/$\\$70$}   & $1.0$  & $1.0$ & $1,573$     & \makecell{$5/$\\$54$}     & $1.0$ & $27.6$  & $-48.2$  \\ \cline{2-11} 
\multirow{1}{*}{}        & $20$                        & $113$     & \makecell{$30/$\\$77$}   & $1.0$  & $1.0$ & $31,457$    & \makecell{$151/$\\$1347$} & $1.0$ & $278.4$ & $92.9$   \\ \hline \hline

\multirow{1}{*}{Tandem}  & $2047$                 & $33$      & \makecell{$1/$\\$41$}    & $0.498$  & $0.498$ & $2,392$     & \makecell{$3/$\\$38$}     & $0.498$ & $72.5$  & $-1.4$   \\ \cline{2-11} 
\multirow{1}{*}{$(c)$}   & $4095$                 & $66$      & \makecell{$1/$\\$141$}   & $0.499$  & $0.499$ & $9,216$     & \makecell{$11/$\\$265$}   & $0.499$ & $139.6$ & $48.7$   \\ \hline 

\end{tabular}
\end{center}
\end{table}

\textbf{Jackson Queuing Network}. A Jackson queuing network consists of $N$ interconnected nodes (queues) with infinite queue capacity. Initially, all queues are considered empty. Each station is connected to a single server which distributes the arrived jobs to different stations. Customers arrive as a Poisson stream with intensity $\lambda$ for $N$ queues. The model is taken from \cite{Jackson57,HahnHWZ09}. 
% A customer, upon completing service at a node $i$, either leaves the network or enters another node $j$. 
% We consider the case with $N=4,5$ with constant $\lambda=5$. 
We compute the probability that, within 10 time units, the first queue has more that $3$ jobs and the second queue has more than $5$ jobs, given by $\prob_{=?} \; [ \; true \; \until^{\leqslant 10} \; (jobs\_1\geqslant 4 \; \land \; jobs\_2\geqslant 6)]$.

Table~\ref{tbl-compare-stamina-infamy} summarizes the results for this model. STAMINA uses roughly equal time to construct and analyze the model for $N=5$, whereas INFAMY takes significantly longer to construct the state space, making it slower in overall runtime. For $N=4$, STAMINA is faster in generating verification results In both configurations, STAMINA only explores approximately one third of the states explored by INFAMY.

\textbf{Cyclic Server Polling System}. % This case study taken from the PRISM's benchmark suite \cite{Kwiatkowska2011,IT90}.
This case study is based on a cyclic server attending $N$ stations. We consider the probability that station one is polled within 10 time units, $\prob_{=?} \; [ \; true \; \until^{\leqslant 10} \; station1\_polled \;]$. Table~\ref{tbl-compare-stamina-infamy} summarizes the verification results for $N=12,16,20$. The probability of station one being polled within $10$ seconds is $1.0$ for all configurations. Similar to previous case studies, STAMINA explores significantly smaller state space. The advantage of STAMINA in terms of runtime starts to manifest as the size of model (and hence the state space size) grows.
% $\stTermToler = 10^{-6}$ is sufficient to generate enough states to obtain accurate probability. As in previous case studies, STAMINA uses fewer states than INFAMY, which results in a significantly lower overall runtime for larger system models with $N=20$.

\textbf{Tandem Queuing Network}. A tandem queuing network is the simplest interconnected queuing network of two finite capacity ($c$) queues with one server each \cite{Kwiatkowska2011}. Customers join the first queue and enter the second queue immediately after completing the service. This paper considers the probability that the first queue becomes full in $0.25$ time units, depicted by the CSL property $\prob_{=?} \; [ \; true \; \until^{\leqslant 0.25} \; queue1\_full \;]$.

As seen in Table~\ref{tbl-compare-stamina-infamy}, there is almost fifty percent probability that the first queue is full in $0.25$ seconds irrespective of the queue capacity. 
% Similar to the polling system case study, STAMINA explores enough states without needing the property-guided state expansion. 
As in the polling server, STAMINA explores significantly smaller state space. The runtime is similar for model with smaller queue capacity ($c=2047$). But the runtime improves as the queue capacity is increased.

%%% Local Variables:
%%% mode: latex
%%% TeX-master: "main"
%%% End:

\section{Conclusions}

This paper presents an infinite-state stochastic model checker,
STAMINA, that uses path probability estimates to generate states with
high probability and truncate unlikely states based on a specified
threshold. Initial state construction is property agnostic, and the
state space is used for stochastic model checking of a given CSL
property. The calculated probability forms a lower and upper bound on the probability for the CSL property, which is guaranteed to include the actual probability. Next, if finer precision of the probability bound is required, it uses a property-guided state expansion technique to explore states to tighten the reported probability range
incrementally. Implementation of STAMINA is built on top of the PRISM
model checker with tight integration to its API. Performance and accuracy evaluation is performed on case studies taken from various application domains, and shows significant improvement over the state-of-art infinite-state stochastic model checker INFAMY. For future work, we plan to investigate methods to determine the reduction factor on-the-fly based on the probability bound. Another direction is to investigate heuristics to further improve the property-guided state expansion, as well as, techniques to dynamically remove unlikely states.

\section*{Acknowledgment}

Chris Myers is supported by the National Science Foundation under  CCF-1748200. Any opinions, findings, and conclusions or recommendations expressed in this material are those of the authors and do not necessarily reflect the views of the NSF.

%%% Local Variables:
%%% mode: latex
%%% TeX-master: "main"
%%% End:

%
\newpage
\bibliographystyle{splncs04}
\bibliography{references}

\end{document}